\documentstyle[preprint,prl,aps]{revtex}
\begin{document}
\draft
\title{Wigner crystal states for the two-dimensional
electron gas in a double quantum well system}
\author{Lian Zheng and H. A. Fertig}
\address{Department of Physics and Astronomy,
University of Kentucky, Lexington, Kentucky 40506}
\maketitle
\begin{abstract}
Using the Hartree-Fock approximation, we calculate the
energy of different Wigner
crystal states for the two-dimensional electron gas
of a double quantum well system in a strong magnetic field.
Our calculation takes interlayer hopping as well as an in-plane magnetic
field into consideration.
The ground
state at small layer separations is a one-component triangular lattice
Wigner state.
As the layer separation is increased, the ground state first undergoes
a transition to two stacked square lattices,
and then undergoes another transition at an even larger
layer separation
to a two-component triangular lattice.
The range of the layer separation at which the two-component
square lattice occurs as the ground state shrinks, and eventually
disappears,
as the interlayer hopping is increased.
An in-plane magnetic field induces another phase transition
from a commensurate to a incommensurate state,
similar to that of $\nu=1$ quantum Hall state
observed recently.
We calculate the critical value of the in-plane field of the transition
and find that
the anisotropy of the Wigner state,
{\it i.e.,}, the relative orientation of the crystal
and the in-plane magnetic field,
has a negligible effect on the critical value for low filling fractions.
The effect of this anisotropy on the low-lying phonon energy is discussed.
A novel exerimental geometry is proposed
in which the parallel magnetic field is used to enhance the
orientational correlations in the ground state when the
crystal is subject to a random potential.

\end{abstract}
\narrowtext
\section{introduction}
An electron gas is expected to condense into a Wigner crystal \cite{wc}(WC)
below some critical density.  This condensation occurs
when the Coulomb energy, which
tends to localize electrons into
individual lattice sites to keep them as far apart
as possible from each other,
dominates over the kinetic energy, which favors
a smooth variation of electron density.
In the absence of a magnetic field, the kinetic energy of a
two-dimensional electron
system (2DES) scales like $K=\hbar^2/m^*a^2$, while the Coulomb energy
scales like $V=e^2/\epsilon a$, where $a$ is the mean inter-electron
distance and $\epsilon$ is the dielectric constant of the host
material, and $m^*$ and $e$ are the electron mass and charge respectively.
The relevant parameter is the ratio $r_s=V/K=a/a_B$, where
$a_B=\hbar^2\epsilon/m^*e^2$ is the Bohr radius.
Monte Carlo simulation \cite{mc}
predicts that a 2DES crystallizes
for $r_s\geq35$.
When a strong magnetic field is applied perpendicular
to the 2DES, the situation is changed qualitatively,
as the kinetic energy is quenched into discreet Landau levels, and
the zero-point fluctuations in the lowest Landau level
are confined within a magnetic length $l_o=(\hbar c/eB_\perp)^{1/2}$,
where $c$ is the speed of light and $B_\perp$ is the applied
magnetic field.
Once $l_o$ is sufficiently small compared to the typical
inter-particle distance
$a$, crystallization occurs. The ratio $l_o/a$
can be characterized by the Landau level filling factor
$\nu=2\pi l^2n$, where $n$ is the density of the 2DES.
Crystallization will occur for
sufficiently small $\nu$
for any given density.
Theoretical estimates \cite{gm}
put the critical filling factor of crystallization
at about $\nu_c\sim 1/6$.
In recent years, especially after the observation
of the reentrant insulating phase around the $\nu=1/5$
quantum Hall state \cite{fifth},
there has been an increasing interest in the study of the WC states in
a 2DES in a strong magnetic field. Many experimental results
\cite{fifth,goldman2,andrei,paalanen,besson,buhmann,clark}
are found
to be in some ways consistent with the assumption of
a pinned WC as the ground state.

Recent advances in material-growth technology allows the
fabrication of high quality double quantum well system (DQWS),
in which two interacting 2DES are separated by a distance comparable
to the mean inter-particle distance within the 2DES. This introduces
a new degree of freedom associated with the third direction.
Electron-electron interactions between the layers have been known
to lead to new quantum Hall states \cite{qh}. It also leads
to increased stability of the WC state or other charge density wave
states \cite{ft}.
Recent experiments on two layer systems in wide quantum wells below
filling factor $\nu=1/2$
have shown insulating behavior\cite{suen} similar to that seen in
single layer systems below $\nu=1/5$.
Because Coulomb interactions can lead to
mixing of the electronic states of the two wells, more complicated
structures of WC states become possible in DQWS.
The goal of this paper is to investigate the evolution of
the ground state among different possible WC configurations,
as the parameters of the DQWS are changed, at a small Landau level
filling factor
where a WC is expected to be the ground state.
This is accomplished by calculating,
in the Hartree-Fock approximation,
the ground state
energy of the different WC states.
Our method of calculation is based on the numerical technique developed
in Ref.\ \cite{cote}, which is valid in the strong-field limit.
We take into consideration inter-layer hopping as well as
an in-plane magnetic field.  We ignore the finite thickness of the
quantum wells and treat the electron gas as ideally two-dimensional.
This zero-thickness approximation is an important simplification
in the case of a tilted field,
since the effect of the in-plane magnetic field can
then be included by adding
a phase factor to the wavefunction of the electrons in one of the wells.

At small layer separations, where the electron-electron correlations
between the wells are almost as important as the correlations within
each individual well, electrons will occupy the symmetric state
of a DQWS
to minimize Coulomb energy and to take advantage of
hopping energy.  In this situation, the DQWS behaves essentially
like a single layer system.  The WC at small filling factor is therefore a
triangular lattice.
As the layer separation increases, the 2DES will seek a state where
intralayer correlations are favored over the interlayer correlation.
This leads to a transition to a truly two-component system
when the intralayer Coulomb energy is more important than the
hopping and interlayer interaction energy combined.
Under this condition, each individual well
forms a WC separately with a lattice constant which is appropriate
for the electron density in its own layer.
The two lattices are shifted relative to one another
to lower the interlayer static Coulomb energy (Hartree energy).
We find that, in the absence of interlayer hopping, the one-component
WC is first transformed to a two-component square lattice WC state,
and then transformed to a two-component triangular lattice,
as the layer separation is increased.
The existence of a square lattice WC at intermediate
separations can be understood
by noticing that the square lattice configurations have
lower interlayer interaction energy
than that of the triangular lattices. This gain may exceed
the difference in intralayer Coulomb
energies between a square lattice and a triangular lattice,
which is known to be small\cite{bm}.
The range of the layer separation at which the two-component
square lattice exists shrinks, and eventually disappears, as
the interlayer hopping is increased,
as a result of the expansion of
the one-component triangular phase.
Our result is  consistent with that of Chan and MacDonald\cite{chan}, where
they treat
electrons classically.

A DQWS in a tilted magnetic field has been studied recently.
A new phase transition, driven by the in-plane component of the field,
was observed for $\nu=1$ quantum Hall state\cite{qhe}, and
was explained using an easy-plane itinerant quantum
ferromagnetism description\cite{qht}.
This phase transition happens as a result of the competition
between the hopping energy
and Coulomb interaction energy. The in-plane magnetic field
twists the interlayer phase coherence of the wavefunctions of electrons
in symmetric state. The result is an increase
in interlayer Coulomb energy. At small in-plane field,
the increase in interlayer Coulomb energy is small compared to
the hopping energy,
so the electrons will stay
in the state
dictated by the tunneling part of the Hamiltonian
in order to  take advantage of the hopping
energy.
As the in-plane field becomes stronger, the hopping energy is
reduced, while the cost in Coulomb  energy continues to rise.
When the in-plane field is raised beyond a critical value,
the DQWS system will undergo a transition to a state in which
electrons give up the
hopping energy in order to restore the interlayer correlations.
The same physics occurs for either a $\nu=1$ quantum Hall state or
a one-component
WC states, since for both cases there is
strong interlayer coherence
before the application of an in-plane field.
Since a WC breaks the rotational symmetry of the system,
the phase transition should, in principle, depend quantitatively on the
angle between the direction of the in-plane field
and the crystal axes.
We have calculated the critical value of the in-plane field and found
that this dependence is very weak, practically unobservable at small
filling factors.

Nevertheless, a parallel magnetic field can have important
consequences, particularly for the long-wavelength physics
of the system.
The broken orientational symmetry of the groundstate due to a
parallel magnetic field implies that there is a restoring force
on the crystal if it is misoriented with the field.
This effect can be very important if the system is subject
to a weak random potential, which destroys long-range orientational
order in this system\cite{cha}.  We propose a novel
experimental geometry, in which the 2DEG is cooled through
its freezing transition in the presence of a parallel
magnetic field.  The force that tends to align the crystal
axes perpendicular to the parallel field should enhance the
orientational order in the groundstate.
It has been shown recently\cite{cha} that the
depinning electric field\cite{goldman2}
of a disordered WC is sensitive to the orientational correlations
in the system, and tends to be reduced as orientational order
increases.  Thus an increase in orientational order can in
principle be detected experimentally through
the transport properties of the system.

This article is organized as follows. In Sec. 2, we describe
the Hartree-Fock approximation for the interacting 2DES
in a DQWS in
a strong magnetic field.
In Sec. 3, we present and discuss the numerical results.
Sec. 4 discusses the effect of a parallel magnetic
field on some of the crystal properties.
A brief summary in Sec. 5 concludes this article.

\section{Hartree-Fock approximation}
In the absence of an in-plane magnetic field, the Hartree-Fock
approximation for a DQWS has been clearly presented
in detail in Ref. \cite{cote}.
We need only to extend it to the case of a tilted magnetic field.
In a Hartree-Fock approximation, one
treats the interacting Hamiltonian as that of
free electrons in the mean-field potential
determined by a given electronic state
and then self-consistently solves for the state.
In our present case, the electronic state being sought is a WC.
The characteristic of
a WC is a periodically modulated charge density.
We therefore choose the Fourier transformed electron density at
corresponding reciprocal lattice vectors (RLV) as the order parameters
of the WC states. Following this idea, we
define $\rho_{ij}({\bf q})=\int d^2re^{-i{\bf q\cdot r}}
\psi^\dagger_i({\bf r})
\psi_j({\bf r})$,
where  $i,j=1,2$, labeling the two layers,
$\psi^\dagger$ (
$\psi_j({\bf r})$)
is electron creation (annihilation) operator.
One obtains, in the lowest Landau level with the Landau gauge,
\begin{equation}
\rho_{ij}({\bf q})={1\over g}\sum_{\alpha\beta}e^{-i
q_x(\alpha+\beta)/2}\delta_{\beta,\alpha+q_y}C^\dagger_{i\alpha}
C_{j\beta},
\label{equ:rho1}
\end{equation}
where $\alpha$ and $\beta$
are the single-particle states of the lowest Landau
level.  In the above expression,
$g=\Omega/2\pi l_o^2$ is the Landau level degeneracy,
where $\Omega$ is the area of the DQWS.
The order parameters have, by definition, the property that
$<\rho_{ij}({\bf q})>=<\rho_{ji}(-{\bf q})>^*$.  We restrict ourselve to
seek only the WC states where the charge distribution in one layer
is the same as that of the other layer, except that one is rigidly
shifted by a displacement of ${\bf a}$ relative to the other, {\it i.e.},
$<\rho_{22}({\bf q})>=e^{i{\bf q}\cdot{\bf a}}<\rho_{11}({\bf q})>$.
This leaves us with only two
independent sets of the order parameters to be obtained.
The order parameters are related to Green's function as
\begin{equation}
<\rho_{ij}({\bf q})>
=G_{ji}({\bf q},\tau=0^-),
\label{equ:rho2}
\end{equation}
where $G_{ij}({\bf q},\tau)=\int d^2re^{-i{\bf q}\cdot{\bf r}}
G_{ij}({\bf r},{\bf r},\tau)$
and $G_{ij}({\bf r},{\bf r^\prime},\tau)=-<T_\tau\psi_i
({\bf r},\tau)\psi_j^\dagger({\bf r}^\prime)>$.
In the lowest Landau level, the Green's functions
become
\begin{equation}
G_{ij}({\bf q},\tau)={1\over g}\sum_{\alpha\ \beta}e^{-iq_x
(\alpha+\beta)/2}\delta_{\alpha,\beta+q_y}G_{ij}(\alpha,\beta,\tau),
\label{equ:g1}
\end{equation}
with
\begin{equation}
G_{ij}(\alpha,\beta,\tau)=-<T_\tau C_{i\alpha}
(\tau)C_{j\beta}^\dagger>.
\label{equ:g2}
\end{equation}

The Green's functions are to be obtained by
self-consistently solving the equation of motion
\begin{equation}
{\partial\over\partial\tau}G_{ij}(\alpha,\beta,\tau)
+\delta(\tau)\delta_{ij}\delta_{\alpha,\beta}
+<T_\tau [H,C_{i\alpha}(\tau)]C^\dagger_{j\beta}>=0.
\label{equ:g3}
\end{equation}

The next step is to approximate the equations of motion
in the Hartree-Fock approach.
We will treat a tilted magnetic field from the beginning.
The result for
a perpendicular field is recovered by simply setting
the in-plane component of the field $B_\parallel=0$.
Under the zero well-thickness approximation, the only effect
of the in-plane magnetic field is to add a phase factor
to the electronic wave function of one of the layers.
Let $k_B=d/l_\parallel^2$, with $d$ the separation between
the wells, and $l_\parallel=(\hbar c/eB_\parallel)^{1/2}$.
If one chooses the direction of $B_\parallel$ as the x-axis,
the single-particle eigenstates become
\begin{eqnarray}
\phi_{1\alpha}({\bf r})&&={1\over\sqrt{l_o\sqrt{\pi\Omega}}}
e^{i\alpha y}e^{-{(x-\alpha l_o^2)^2/2l_o^2}}\nonumber \\
\phi_{2\alpha}({\bf r})&&=e^{ik_B y}\phi_{1\alpha}({\bf r}).
\label{equ:wf1}
\end{eqnarray}

The Hamiltonian of the DQWS contains kinetic energy,
(which is a trivial constant for the lowest Landau level,) interlayer
hopping energy, and the Coulomb interaction energy.

\begin{eqnarray}
H=&&-\sum_{i\alpha}\mu C^\dagger_{i\alpha}C_{i\alpha}-
te^{-k_B^2l_o^2/4}\sum_{\alpha}
(C^\dagger_{1\alpha}C_{2\alpha-k_B}+
C^\dagger_{2\alpha-k_B}C_{1\alpha}) \nonumber \\
&+&{1\over2}\sum_{ij}\sum_{\alpha\alpha^\prime\beta\beta^\prime}
V_{ij}(\alpha,\alpha^\prime,\beta,\beta^\prime)
C^\dagger_{i\alpha}C^\dagger_{j\beta}C_{j\beta^\prime}C_{i\alpha^\prime},
\label{equ:h1}
\end{eqnarray}
where $\mu$ is chemical potential
to be fixed at the end of calculation for a given electron density,
and $V_{ij}(\alpha,\alpha^\prime,\beta,\beta^\prime)$
is the matrix element of Coulomb potential.
The hopping parameter $t$ is suppressed in the presence
of an in-plane field by a factor $e^{-k_B^2l_o^2/4}$,
which comes from the matrix element $<1\alpha|t|2\alpha^\prime>=
te^{-k_B^2l_o^2/4}\delta_{\alpha,\alpha^\prime+k_B}$.
Physically, this means that the electrons tunnel along the
direction of the total magnetic field.
Performing the Hartree-Fock pairing
$C^\dagger_1C^\dagger_2C_{2^\prime}C_{1^\prime}
\approx<C^\dagger_1C_{1^\prime}>
C^\dagger_2C_{2^\prime}-
<C^\dagger_1C_{2^\prime}>C^\dagger_2C_{1^\prime}$ on Eq.\ (\ref{equ:h1}),
one obtains
\begin{equation}
H=-g\sum_i\mu\rho_{ii}(0)-gt[\rho_{21}({\bf k}_B)+\rho_{12}(-{\bf k}_B)]
+g{e^2\over \epsilon l_o}
\sum_{\bf q}\sum_{ij}U_{ij}({\bf q})\rho_{ji}({\bf q}),
\label{equ:h2}
\end{equation}
where ${\bf k}_B=k_B\hat{\bf y}$ and
\begin{eqnarray}
U_{11}({\bf q})&&=[V_a({\bf q})-V_b({\bf q})]<\rho_{11}(-{\bf q})>
+V_c({\bf q})<\rho_{22}(-{\bf q})>,\nonumber \\
U_{12}({\bf q})&&=-V_d({\bf q})<\rho_{21}(-{\bf q})>,
\label{equ:uij}
\end{eqnarray}
with $U_{21}$ and $U_{22}$ obtained by interchanging the indices
$1$ and $2$. In the above expressions, $V_a$, $V_b$, $V_c$, and $V_d$ are
the direct and
exchange terms for the interlayer and intralayer
Coulomb interactions
\begin{eqnarray}
V_a(q)&=&{1\over ql_o}e^{-q^2l_o^2/2}
(1-\delta_{{\bf q},0})\nonumber \\
V_b(q)&=&\sqrt{\pi\over2}e^{-q^2l_o^2/4}I_0(q^2l_o^2/4)
\nonumber \\
V_c(q)&=&e^{-qd}V_a(q)\nonumber \\
V_d(q)&=&\int_0^\infty dxJ_0(xql_o)e^{-x^2/2-xd/l_o}
\label{equ:va},
\end{eqnarray}
where $J_0$ and $I_0$ are the zero-th order
Bessel function and modified Bessel function,
respectively.
It is worthwhile to notice that since Coulomb scattering
is not expected to move electrons from one well
to the other, the matrix element
$V_{ij}(\alpha,\alpha^\prime,\beta,\beta^\prime)$
is unchanged by the presence of an in-plane field,
{\it i.e.,} Eq.\ (\ref{equ:uij}) and Eq.\ (\ref{equ:va}) are
exactly the same as the expressions for $B_\parallel=0$.
This means that
the commutator $[H_c,C_{i\alpha}]$,
where $H_c$
is the Coulomb interactions part of
the Hartree-Fock Hamiltonian of Eq.\ (\ref{equ:h2}),
is unchanged by the presence of the in-plane field.
Letting $A_{ij}(\alpha,\beta)=
-<T_\tau[H_c,C_{i\alpha}](\tau)C^\dagger_{j\beta}>$,
we have
\begin{equation}
A_{ij}({\bf q},\tau)=-{e^2\over\epsilon l_o}
\sum_{k{\bf p}}U_{ik}({\bf p}-{\bf q})e^{{il_o^2
}{\bf q}\wedge{\bf p}/2}
G_{kj}({\bf p},\tau),
\label{equ:a1}
\end{equation}
where ${\bf q}\wedge{\bf p}=q_xp_y-q_yp_x$.
The effect of the in-plane field is contained explicitly in
$H_t$, the hopping term in Eq.\ (\ref{equ:h2}).
Denoting $F_{ij}(\alpha,\beta)=
-<T_\tau[H_t,C_{i\alpha}](\tau)C^\dagger_{j\beta}>$,
one obtains
\begin{eqnarray}
&&F_{11}({\bf q},\tau)=te^{-k_B^2l_o^2/4}
e^{-{il_o^2}q_xk_B/2}
G_{21}({\bf q}-{\bf k}_B,\tau) \nonumber \\
&&F_{12}({\bf q}+{\bf k}_B,\tau)=te^{-k_B^2l_o^2/4}
e^{-{il_o^2}q_xk_B/2}
G_{22}({\bf q},\tau) \nonumber \\
&&F_{21}({\bf q}-{\bf k}_B,\tau)=te^{-k_B^2l_o^2/4}
e^{{il_o^2}q_xk_B/2}
G_{11}({\bf q},\tau) \nonumber \\
&&F_{22}({\bf q},\tau)=te^{-k_B^2l_o^2/4}
e^{{il_o^2}q_xk_B/2}
G_{12}({\bf q}+{\bf k}_B,\tau)
\label{equ:f1}
\end{eqnarray}

One can see, from the above expressions, that
in the presence of an in-plane field the hopping
Hamiltonian attempts to
shift the positions of
the non-vanishing interlayer order parameters $<\rho_{12}>$
and $<\rho_{21}>$
in reciprocal space from ${\bf q}={\bf G}$ to
${\bf q}={\bf G}\pm{\bf k}_B$.
This is  merely a reflection of the fact
that the charge distributions in the two layers will
be relatively shifted, because the electrons intend to
tunnel along the direction of the total magnetic field.
If the electrons of the DQWS are in the symmetric
state for $B_\parallel=0$,
where electron distributions in the two layers
are directly on top of each other, the inclusion of an
in-plane field damages the original interlayer correlations
and results in an increase in the interlayer Coulomb energy.
In a pseudo-spin description\cite{qht} of the DQWS,
the tunneling
behaves like a tumbling magnetic field,
which twists
the interlayer phase coherence of the symmetric states.
In an attempt to minimize the total energy, a DQWS
is forced to choose between the hopping energy and
the Coulomb energy.
The electrons can only take advantage of the hopping energy at the
cost of increasing the interlayer Coulomb interaction.

Putting Eq.\ (\ref{equ:a1}) and
Eq.\ (\ref{equ:f1}) into Eq.\ (\ref{equ:g3}), one obtains
the desired equations of motion for $G_{11}$ and $G_{21}$
in Matsubara frequencies
\begin{eqnarray}
-{e^2\over\epsilon l_o}\sum_{{\bf q}^\prime}
[e^{{il_o^2{\bf q}\wedge{\bf q}^\prime/2}}
\widetilde U^*_{11}({\bf q}-{\bf q}^\prime)
\widetilde G_{11}({\bf q}^\prime,i\omega_n)
+e^{{il_o^2{\bf q}\wedge{\bf q}^{\prime-}/2}}
\widetilde U^*_{21}({\bf q}-{\bf q}^{\prime-})
\widetilde G_{21}({\bf q}^{\prime-},i\omega_n)]
&&\nonumber \\
+(i\hbar\omega_n+\mu)\widetilde G_{11}({\bf q},i\omega_n)
+te^{-{k_B^2l_o^2/4}}e^{-{il_o^2q_xk_B/2}}
\widetilde G_{21}({\bf q}
-{\bf k}_B,i\omega_n)=\hbar\delta_{{\bf q},0},&& \nonumber \\
-{e^2\over\epsilon l_o}\sum_{{\bf q}^\prime}
[e^{{il_o^2{\bf q}^-\wedge{\bf q}^\prime/2}}
\widetilde U_{21}({\bf q}^\prime-{\bf q}^-)
\widetilde G_{11}({\bf q}^\prime,i\omega_n)
+e^{{il_o^2{\bf q}^-\wedge{\bf q}^{\prime-}/2}}
\widetilde U_{11}({\bf q}-{\bf q}^\prime)
\widetilde G_{21}({\bf q}^{\prime-},i\omega_n)]
&&\nonumber \\
+(i\hbar\omega_n+\mu)\widetilde G_{21}({\bf q}^-,i\omega_n)
+te^{-{k_B^2l_o^2/4}}e^{il_o^2q_xk_B/2}
\widetilde G_{11}({\bf q}^-+{\bf k}_B,i\omega_n)=0,&&
\label{equ:g4}
\end{eqnarray}
where we have adopted the following notations\cite{cote}
\begin{eqnarray}
&&\widetilde U_{ij}({\bf q})=e^{i{\bf q}\cdot {\bf a}/2}U_{ij}({\bf q})
\nonumber \\
&&\widetilde G_{ij}({\bf q})=e^{-i{\bf q}\cdot {\bf a}/2}G_{ij}({\bf q})
\nonumber \\
&&<\widetilde\rho_{ij}({\bf q})>
=e^{-i{\bf q}\cdot {\bf a}/2}<\rho_{ij}({\bf q})>
\nonumber \\
&&{\bf q}^{\pm}={\bf q}\pm {\bf k}_B\ \ \ {\rm or}\ \ \ {\bf q}.
\label{equ:em1}
\end{eqnarray}

Eq. (\ref{equ:g4}) and Eq.\ (\ref{equ:em1}) are intended
to be applicable to different possible WC states.
We need to set correct conditions for each situation.
For the case of a perpendicular magnetic field, one has
${\bf k}_B=0$, ${\bf q}^{\pm}={\bf q}$, and ${\bf a}$ as
the relative shift of the charge distributions between the layers.
As mentioned earlier, there are two possible
phases in the presence of an in-plane field,
electrons can choose either to take advantage of
the hopping energy at the cost of increased Coulomb energy, or
to maintain good correlations for reducing the Coulomb energy
at the cost of giving up the hopping energy. For the former
case, one has ${\bf q}^{\pm}={\bf q}\pm {\bf k}_B$, and ${\bf a}=
k_B l_o^2\hat{\bf x}$. For the latter case, one has
${\bf q}^{\pm}={\bf q}$ and can effectively set $t=0$
(see Eq.\ (\ref{equ:h2}).)

Eq.\ (\ref{equ:em1}) can be rearranged into following compact
matrix form,
which is convenient for numerical evaluation,
\begin{equation}
[(i\hbar\omega_n+\mu)-D]\widetilde G=\hbar B
\label{equ:em2}
\end{equation}
where
\begin{eqnarray}
&&B=[1,0,0,0,0,.....]\nonumber \\
&&\widetilde G=[\widetilde G_{11}({\bf q}_1),\widetilde G_{21}({\bf q}_1^-),
\widetilde G_{11}({\bf q}_2),\widetilde G_{21}({\bf q}_2^-),...]
\label{equ:bg1}
\end{eqnarray}
with $\{{\bf q}_1,{\bf q}_2,{\bf q}_3,...\}$ arranged in the order
of increasing magnitude. The non-zero elements
of the Hermitian matrix $D$ are
\begin{eqnarray}
&&D_{2i-1,2i}=-te^{-{k_B^2l_o^2/4}}
e^{-il_o^2q_xk_B/2}\nonumber \\
&&D_{2i-1,2j-1}={e^2\over\epsilon l_o}e^{il_o^2{\bf q}_i
\wedge{\bf q}_j/2}\widetilde U^*_{11}({\bf q}_i-{\bf q}_j)
\nonumber \\
&&D_{2i-1,2j}={e^2\over\epsilon l_o}e^{il_o^2{\bf q}_i
\wedge{\bf q}^-_j/2}\widetilde U^*_{21}({\bf q}_i-{\bf q}^-_j)
\nonumber \\
&&D_{2i,2i-1}=-te^{-{k_B^2l_o^2/4}}
e^{il_o^2q_xk_B/2}\nonumber \\
&&D_{2i,2j-1}={e^2\over\epsilon l_o}e^{il_o^2{\bf q}^-_i
\wedge{\bf q}_j/2}\widetilde U_{21}({\bf q}_j-{\bf q}^-_i)
\nonumber \\
&&D_{2i,2j}={e^2\over\epsilon l_o}e^{il_o^2{\bf q}^-_i
\wedge{\bf q}^-_j/2}\widetilde U_{11}({\bf q}_i-{\bf q}_j)
\label{equ:d1}
\end{eqnarray}

Eq.\ (\ref{equ:em2}) can be solved by diagonalizing the matrix $D$.
If $V_k$ and $\omega_k$ are respectively the $k$-th eigenvector
and eigenvalue of $D$, we obtain \cite{cote}
\begin{eqnarray}
&&<\widetilde\rho_{11}({\bf q}_i)>=\sum_k^{k_{max}}V_k(2i-1)V_k(1)
\nonumber \\
&&<\widetilde\rho_{12}({\bf q}^-_i)>=\sum_k^{k_{max}}V_k(2i)V_k(1)
\label{equ:s1}
\end{eqnarray}
where $k_{max}$ is determined by fixing the chemical potential
\begin{equation}
<\rho_{11}(0)>=\nu/2.
\label{equ:r11}
\end{equation}

With the order parameters known, we can obtain
the ground state energy per electron,
which will be shown in the next section, from Eq.\ (\ref{equ:h2}),
\begin{eqnarray}
\varepsilon=&&-{1\over\nu}te^{-{k_B^2l_o^2/4}}
[<\rho_{21}({\bf k}_B)>+<\rho_{12}(-{\bf k}_B)>]
\nonumber \\
&&-{e^2\over\nu\epsilon l_o}\sum_{\bf q}\{
V_d( q^-)|<\rho_{12}({\bf q}^-)>|^2 \nonumber \\
&&-[V_a(q)-V_b(q)
+V_c(q){\rm cos}({\bf q}\cdot{\bf a})]|<\rho_{11}({\bf q})>|^2\}.
\label{equ:e1}
\end{eqnarray}

\section{numerical results and discussions}
We now discuss our numerical results for the
different WC states. For this discussion, we
compare the energy of the
different WC states and then find the phase
diagrams as the sample parameters are changed.
The order parameters $<\rho_{ij}({\bf q})>$
are obtained from a numerical analysis on Eq.\ (\ref{equ:em2}),
in which well convergent results are obtained by keeping
16 (24) shells in reciprocal-lattice vectors for triangular (square)
WC states. In the following, we will first discuss the situation
with a perpendicular field and then the situation with a tilted
field.

There are several possible configurations for a WC
in a strong perpendicular field for
different layer separations
and hopping strength.
At small layer separations, the electronic states of the different wells
mix to form symmetric and anti-symmetric states. In the ideal case
of $d=0$, all electrons reside in the symmetric state for any value
of hopping.
A DQWS under this condition behaves as a single layer system.
A WC in the symmetric state has
a lattice constant $a_o$ such that $(\sqrt3/2)a_o^2=2\pi l_o^2/\nu$.
The order parameters at corresponding RLVs are
\begin{eqnarray}
<\rho_{ss}({\bf G})>&&={1\over g}\sum_{\alpha \beta}
e^{-iG_x(\alpha+\beta)/2}\delta_{\beta,\alpha+G_y}
<C_{s\alpha}^\dagger
C_{s\beta}>\not=0,\nonumber \\
<\rho_{ij}({\bf G})>&&={1\over2}<\rho_{ss}({\bf G})>
\ \ \ {\rm for}\ \ \ i,j=1,2,
\label{equ:ss1}
\end{eqnarray}
where $C_{s(a)}=(1/\sqrt2)(C_1\pm C_2)$. The above expression
also shows that the charge distributions for the two layers
are directly on top of each other,{\it i.e.,} ${\bf a}=0$.
At finite, but small,
layer separations, the WC state of a DQWS is essentially
the above one-component triangular (OCT) lattice.
For large enough layer separations, the symmetric state is
no longer energetically favored, as the system begins to
seek a state where electrons within the same layer are
more strongly correlated than electrons in different layers.
In the large $d$ limit,
a DQWS becomes two independent single layers.
Electrons in each well form their
own triangular WC. These two-component triangular (TCT) lattices
have a lattice constant $(\sqrt3/2)a_o^2=2\pi l_o^2/(\nu/2)$, larger
by a factor of $\sqrt2$ than that of the OCT
lattice discussed above.
To minimize static interlayer Coulomb energy, the two WCs are
relatively shifted so that the lattice sites of one WC
are at the centers of the triangles of the other WC lattice,
{\i.e.,} ${\bf a }=(1/3)({\bf a}_o+{\bf b}_o)$, where ${\bf a}_o$
and ${\bf b}_o$ are the primary lattice vectors.
A shifted two-component square (TCS) lattice WC state
with ${\bf a }=(1/2)({\bf a}_o+{\bf b}_o)$, where ${\bf a}_o$
and ${\bf b}_o$ are the primary lattice constant of the square
WC, has lower interlayer Coulomb energy than the above
TCT WC state,
since the lateral distance between
a electron in one layer and its nearest electron in the
other layer in a shifted TCS lattice structure
is larger than that in a shifted TCT lattice structure.
For an intermediate range of layer separations,
this TCS lattice structure may become the ground state
of a DQWS.

In Fig.\ (\ref{fig1}), we show the energies of the three
different WC structures discussed above as functions of
layer separation $d$ for different value of hopping $t$.
The lowest energy states are the OCT
WC at small $d$, the TCS
WC at intermediate $d$, and
the TCT WC at large $d$.
The range of layer separations at which the TCS
WC exists as the ground state shrinks, and finally
disappears, when $t$ is increased, as a result
of the expansion of the OCT WC phase.
The important conclusion from Fig.\ (\ref{fig1}) is
that for weak interlayer hopping,
two structural phase transition should be
expected when the layer separation is increased:
first from a OCT WC to a TCS WC,
then from a TCS WC to a TCT WC.

Next, we consider the influence from an in-plane magnetic field.
Our discussion will concentrate on the situation
where the system was in the symmetric state prior to
the application of the in-plane field, since the effects
of an in-plane field are the largest in the symmetric state
due to the
existence of strong inter-layer coherence.
As a result of the competition between hopping energy and Coulomb energy,
a DQWS in a tilted field can be in either of the two different
ground states depending on the value of the
in-plane field.  One is the symmetric state WC (SSWC) described by
Eq.\ (\ref{equ:ss1}), where ${\bf a}=0$ and the electrons
feel no effect of the hopping and $B_\parallel$
has no effect.
The other,
which is generally relevant
at a small in-plane field,
is the state in which
electrons
form linear combinations of the states in the two wells that
are displaced by the total magnetic field.
We call this state a twisted symmetric state.
Defining
\begin{equation}
{\cal C}_{s(a)\alpha}={1\over2}[C_{1\alpha}\pm C_{2\alpha-k_B}],
\label{equ:cas1}
\end{equation}
the order parameters for a twisted symmetric state WC
(TSSWC) are
\begin{equation}
<\varrho_{ss}({\bf G})>=
{1\over g}\sum_{\alpha \beta} e^{-iG_x(\alpha+\beta)/2}
\delta_{\beta,\alpha+G_y}<{\cal C}^\dagger_{s\alpha}
{\cal C}_{s\beta}>\not=0.
\label{equ:tswc1}
\end{equation}

The order parameters in layer representation
$<\rho_{ij}({\bf G})>$ for this
TSSWC can be obtained from the above expression by making
use of the operator
relation of Eq.\ (\ref{equ:cas1}),
\begin{eqnarray}
&&<\rho_{11}({\bf G})>={1\over2}<\varrho_{ss}({\bf G})>\nonumber \\
&&<\rho_{22}({\bf G})>={1\over2}e^{iG_xk_Bl_o^2}<\varrho_{ss}({\bf G})>
\nonumber \\
&&<\rho_{12}({\bf G}-{\bf k}_B)>
={1\over2}e^{iG_xk_Bl_o^2}<\varrho_{ss}({\bf G})>
\nonumber \\
&&<\rho_{21}({\bf G}+{\bf k}_B)>
={1\over2}e^{iG_xk_Bl_o^2}<\varrho_{ss}({\bf G})>.
\label{equ:tswc2}
\end{eqnarray}

 From the above expression, one can see the obvious effects
of the in-plane field: it shifts
the charge distributions of the two layers relatively
by an amount ${\bf a}=k_Bl_o^2\hat{\bf x}=d(B_\parallel/B_\perp)
\hat{\bf x}$;
The positions of the non-zero interlayer
order parameters $<\rho_{12(21)}({\bf q})>$
in the reciprocal vector space are shifted from
${\bf G}$ to ${\bf G}^\pm={\bf G}\pm {\bf k}_B$.
The result is an increase in interlayer Coulomb energy.
At small values of in-plane field, the increase in the Coulomb
energy is small and can be compensated by the hopping energy,
so the TSSWC is favored over the SSWC. As the
in-plane field increases, the cost in Coulomb energy increases
while the hopping energy decreases.
When the gain in the hopping energy can no longer
compensate the cost in Coulomb energy for a strong enough in-plane
field, the DQWS undergoes a transition from the TSSWC
to the SSWC, where the total energy of the system
is lowered by giving up the hopping energy to
restore the original interlayer coherence.
In Fig.\ (\ref{fig2}), we show the energies of the TSSWC and
SSWC as functions of the in-plane field for different
values of hopping energy. It is clear, from the figure,
that the TSSWC is the energetically favored ground state
at small in-plane field while the SSWC is the energetically
favored ground state at large in-plane fields.
The critical value of the in-plane field for this phase
transition as
a function of the hopping energy is
shown in Fig.\ (\ref{fig3}),
larger critical values of $B_\parallel/B_\perp$ for
larger values of hopping $t$.

Since a WC breaks the rotational symmetry of the system,
the properties of a DQWS, in principle, depend on
the angle between the in-plane field and the crystal axes.
As the WC is pinned by the presence of weak disorder,
the angle can be changed by simply sweeping
the direction of the in-plane field. This
provides a potential opportunity to probe
the orientational order of a WC and to find a unambiguous
signature of the existence of a WC.
In Fig.\ (\ref{fig2}), we show, for $t=0.02,\ t=0.08$, and
$t=0.4$, the energies of the TSSWC for both
the cases where the in-plane field is perpendicular to (the solid-lines),
or parallel with (the dot-lines), one of the crystal axes.
We can see that the differences are small.  The change in the
value of critical in-plane field from the different orientations
of the field is practically indistinguishable.
This is mainly because that the phase transition occurs at
an in-plane field where $|{\bf a}|$ is small compared to
the lattice constant for reasonable sample parameters.
However, we should not rule out the possibility that some
other quantities may have a measurable dependence on the anisotropy
of the WC.
For example, the energy difference of the TSSWC for the two different
orientations of the in-plane field right before the phase transition
is $\Delta\varepsilon\sim 7\times
10^{-4} (e^2/\epsilon l_o)$
for $t=0.08 (e^2/\epsilon l_o)$, which is
on the order of $70mK$ for $e^2/\epsilon l_o\sim 100K$.
{\it IF} the same order of anisotropy  exists in the low-lying phonon modes
(see the next section),
we may have a measurable difference in the anisotropy of the
specific heat at low enough temperatures, say, $T<70mK$.

\section{crystal properties in a parallel magnetic field}
In this section, we discuss some consequences of immersing the
double well system in a parallel magnetic field, when the
electron system is in the TSSWC -- i.e.,
a lattice of particles in symmetric states of the two wells,
with the position of the single-particle orbits in one well
displaced in the direction of the {\it total}
applied magnetic field.  As was discussed in the previous section,
and as shown in Fig. 2, the energy per particle in the electron
lattice depends (weakly) on the orientation of the
crystal axes relative to the parallel magnetic field.
One may think of this effect as an energy cost for having
the bond angles of the lattice deviate from some preferred
direction.  While the energy cost per particle may be small,
long wavelength fluctuations of the lattice
-- where many bond angles deviate from the preferred direction --
will be strongly affected by this energy cost.  This will have
important effects on the long wavelength collective modes
of the system, as well as the state of the crystal in the presence
of a slowly varying random potential (which typically
arises in real heterostructure environments.)

To model the bond-angle energy, we use a continuum elasticity
theory approach.  The energy of a two-dimensional
lattice deformed from a perfect crystal configuration
by a displacement field $u(\vec{r})$ may be written as\cite{landau}
$$
E_0= {1 \over 2} \int {{d^2r} \over {a_0^2}}
\bigl(2\mu u_{ij}^2 + \lambda u_{kk}^2
\bigr),
$$
where $\mu$ and $\lambda$ are Lam\'e coefficients, $a_0$
is the lattice constant,
$u_{ij}=\partial_i u_j+ \partial_j u_i$ is the strain
tensor, and repeated indices are summed over.  The bond angle
field may be written in terms of the displacements in the form
$$
\theta(\vec{r}) \equiv {1 \over 2} \bigl(\partial_x u_y- \partial_y u_x
\bigr),
$$
so that it is natural to write the energy of a configuration in the
presence of parallel magnetic field in the form $E=E_0+E_{\theta}$,
with
$$E_{\theta} \equiv {1\over 2} \epsilon_0 \int d^2r \theta(\vec{r})^2,$$
where $\epsilon_0$ is a phenomenological parameter describing the
energy cost to misalign the crystal with the parallel magnetic
field\cite{c1}.

This model has been studied in the context of a two-dimensional crystal
adsorbed on a periodic substrate\cite{nelson}, where it was noted
that the extra term tends to increase the stability of the crystal
with respect to finite temperature.  It is also interesting to
note that such a bond-angle term increases the stability of
the crystal with respect to quenched disorder.  To see this,
consider a weak random potential acting on the WC,
which has a
long orientational correlation length  $\xi$ (or even quasilong-range
orientational order\cite{cha}).  The energy cost to misalign
a correlated region with the magnetic field scales as $\xi^2$
due to $E_{\theta}$,
whereas the energy gained by aligning with the random potential
scales only as $\sqrt{N_c} \sim \xi$, where $N_c$ is the average number
of electrons in a correlated region.  Thus, for weak enough disorder,
where $\xi$ is large, one expects the correlated regions to align
with the preferred orientational axis.

An interesting possible method to demonstrate this would be to
anneal a WC in the double well system in the presence of a parallel
magnetic field.  It has been shown recently\cite{cha} that a WC in the
presence of a slowly varying random potential freezes
into a state with at best power-law (i.e., quasilong-range)
orientational order.  By cooling the system in the presence of
the parallel field, a preferred orientational axis is picked out
for the crystal, leading to the possibility that one could
obtain a system with true long-range orientational order.
Observing that phenomenon will be possible if the orientational
correlation length is large enough at the freezing
transition that the $E_{\theta}$ term overcomes thermal
fluctuations in the orientation of a correlated region of
the lattice; i.e.,  $\epsilon_0 N_c > T_M$, where $T_M$ is the melting
temperature
of the crystal.  Experimentally, one could probe this effect by
measuring the depinning electric field of the lattice, which has
been shown to be sensitive to orientational correlations of the
crystal\cite{cha}; one expects to see a decrease in the
pinning field if the orientational correlations are increased.
The possibility of creating a WC with long-range order using
parallel magnetic fields in a double well system is currently
under investigation\cite{cha2}.

It is also interesting to consider the effect of the bond-bending
term on the collective mode spectrum of the WC.  It is well known\cite{fuku}
that in a perpendicular magnetic field, the WC supports a phonon mode
dispersing as $\omega(q) \propto q^{3/2}$, and there have been
attempts to measure this directly using rf absorption\cite{andrei}.
Since the bond-angle term $E_{\theta}$ represents a restoring
force on long-wavelength fluctuations, we expect the phonon energy
to change in the present condition.
For this purpose, we write down the appropriate form for the
energy of a crystal deformation in terms of a dynamical matrix
that yields the energy $E=E_0+E_{\theta}$ in the long wavelength
limit.
The low frequency
collective mode frequencies are given by $\lambda_{\alpha}/\omega_c$,
where $\lambda_{\alpha}$ are the eigenvalues of the matrix
$\sigma_y D$, $\sigma_y$ is the Pauli spin matrix, and
$D=D^0+D^{\theta}$ is the total dynamical matrix\cite{cote2}.
$D^0$, the dynamical matrix in the
absence of a parallel magnetic field,
has been shown\cite{bm} to be
$$
D^0_{ij} =b
{{q_iq_j} \over q}
+\sum_{\alpha \beta}A_{ij\alpha\beta}q_\alpha q_\beta,
$$
{\noindent where $b$ and $A_{ij\alpha\beta}$ are constants.
The dynamical matrix associated with
$E_{\theta}$ is easily obtained,}
$$D^{\theta}_{xx} = {{\epsilon_0} \over {4m}} q_y^2,~~~~~
D^{\theta}_{yy} = {{\epsilon_0} \over {4m}} q_x^2,~~~~~
D^{\theta}_{xy}  = D^{\theta}_{xy} =-{{\epsilon_0} \over {4m}} q_xq_y.
$$
By noticing the similarity between $D^{\theta}$ and the second term
in $D^0$, it is easy to show that under the
present condition, the low-lying
phone mode still disperses like $\omega(q)=Cq^{3/2}$, but
the coefficient $C$ is
increased,
with a new contribution
from the dynamic matrix $D^{\theta}$.

\section{conclusion}
Working in the Hartree-Fock approximation, we calculated
the energy of different Wigner states of the two-dimensional electron gas
for a double quantum well system in a strong magnetic field.
We found the phase diagram for the evolution of the WC states in
a strong perpendicular magnetic field when layer separation and
hopping
are changed.
In the absence of interlayer hopping, the ground
state at small layer separations is a one-component triangular lattice
Wigner state which possesses interlayer coherence.
As the layer separation is increased, the ground state first undergoes
a transition to a two-component square lattice
Wigner state,
and then undergoes another transition
at an even larger layer separation
to a two-component triangular lattice Wigner state.
The range of the layer separations at which the two-component
square lattice occurs as the ground state shrinks, and eventually
disappears,
as the interlayer hopping is increased.
We also studied the in-plane magnetic field induced phase transition in the
Wigner state, which has
so far only been studied experimentally for $\nu=1$ quantum Hall state.
We calculated the critical value of the in-plane field for the transition.
We find that the anisotropy of the Wigner state,
{\it i.e.,}, the orientation of the crystal
with respect to the direction of the in-plane magnetic field,
has a negligible effect on
the value of the critical in-plane magnetic field for small
filling factors.
The effect of this anisotropy on the low-lying
phonon energy is discussed.
A possible experimental arrangement for observing
the in-plane field enhancement of the orientational order in the crystal
in the presence of a weak disorder potential
is also discussed.
\section{acknowledgement}
The authors wish to thank Prof. A. H. MacDonald, Dr. Anthony Chan,
and M. C. Cha
for helpful discussions. This work is supported by the National
Science
Foundation through grant No. DMR-9202255.
H.A.F. acknowledges the support of the A.P. Sloan Foundation
and the Research Corporation.

\begin{figure}
\caption{ Energy per particle for a one-component triangular
lattice (OCT), a two-component square lattice (TCS), and a
two-component triangular lattice (TCT) Wigner crystal
as functions of
layer separation $d$ for $\nu=1/4$
at different values of hopping $t$.
(a) $t$=0; (b) $t=0.01(e^2/\epsilon l_o)$;
(c) $t=0.02(e^2/\epsilon l_o)$.}
\label{fig1}
\end{figure}

\begin{figure}
\caption{Energies per particle of a
twisted symmetric state Wigner crystal (TSSWC) and
a symmetric state Wigner crystal
(SSWC) as functions of in-plane field
at $\nu=1/4$ and $d=l_o$ for different
value of hopping. The dash-dot line is for
the SSWC. The solid lines are for the TSSWC with
the in-plane filed perpendicular to one crystal axis.
The dot-lines at $t=0.02, 0.08, 0.4 (e^2/\epsilon l_o)$ are for
the TSSWC with the in-plane field parallel with one
crystal axis.}
\label{fig2}
\end{figure}

\begin{figure}
\caption{
The critical value of $B_\parallel/B_\perp$
as a function of the hopping for $\nu=1/4$ and $d=l_o$.}
\label{fig3}
\end{figure}


\begin{references}
\bibitem{c1} The bond-bending
coefficient $\epsilon_0$ may be estimated for a given set of
parameters by comparing the energies for the two orientations
in Fig. 2, and dividing by $\pi/6$.  For most experimentally
relevant geometries, one finds $\epsilon_0 \ll \mu, \lambda$;
nevertheless, the bond angle term has very important consequences
for the long wavelength physics of this system.
\end{references}
\end{document}